\documentclass[prl,twocolumn,showpacs]{revtex4}

\usepackage{graphicx}

\begin{document}

\title{Polarization and readout of coupled single spins in diamond}
\author{R. Hanson, F. M. Mendoza, R. J. Epstein, and D. D. Awschalom}
\affiliation{Center for Spintronics and Quantum Computation, University of California, Santa Barbara, CA 93106, USA}

\date{\today}

\begin{abstract}
We study the coupling of a single nitrogen-vacancy center in diamond to a nearby single nitrogen defect at room temperature. The magnetic dipolar coupling leads to a splitting in the electron spin resonance frequency of the N-V center, allowing readout of the state of a single nitrogen electron spin. At magnetic fields where the spin splitting of the two centers is the same we observe a strong polarization of the nitrogen electron spin. The amount of polarization can be controlled by the optical excitation power. We combine the polarization and the readout in time-resolved pump-probe measurements to determine the spin relaxation time of a single nitrogen electron spin. Finally, we discuss indications for hyperfine-induced polarization of the nitrogen nuclear spin.
\end{abstract}


\maketitle

The nitrogen-vacancy (N-V) color center in diamond is well suited for studying electronic and nuclear spin phenomena, since its spin can be both initialized and read out via a highly stable optical transition~\cite{Davies76,GruberScience97}. The N-V center has been used as a single photon source~\cite{KurtsieferPRL00} and Rabi oscillations~\cite{JelezkoRabiPRL04} and single-shot readout~\cite{JelezkoAPL02} of the N-V center spin have been demonstrated. Quantum information processing with N-V centers can be envisioned, as ensemble measurements have shown room-temperature electron spin coherence times up to 58~$\mu$s~\cite{KennedyAPL2003}. A two-qubit gate using the nuclear spin of a nearby $^{13}$C atom has already been demonstrated~\cite{{Jelezko2qubitPRL2004}}. However, such an electron-nuclear qubit system is hard to scale to more qubits, and coupling of N-V center spins to each other or to other defect electron spins therefore remains a challenge.

Recently, resonant coupling of single N-V centers to electron spins of substitional nitrogen atoms was detected~\cite{EpsteinNatPhys05}. These measurements have highlighted avenues for extending the physics of the N-V center beyond that of a single electron spin. In a first step, two-spin dynamics can be investigated if an N-V center is strongly coupled to a single substitional nitrogen atom~\cite{WrachtrupCoupling}.

Here, we investigate single N-V centers that are surrounded by a bath of nitrogen spins, but are strongly coupled to only a single nitrogen spin. The magnetic dipolar coupling leads to a splitting in the ESR line of the N-V center, allowing readout of the state of the nitrogen spin. We find that the nitrogen spin can be strongly polarized at magnetic fields where the energy splitting of the nitrogen doublet equals that between the lowest spin levels of the N-V center. The amount of polarization can be controlled by the optical excitation power. We combine the initialization and the readout in time-resolved pump-probe measurements to determine the spin relaxation time of the nitrogen spin. Finally, we discuss indications in the data for polarization of the nitrogen nuclear spin. The polarization and readout of coupled spin states demonstrated in this work outline the future possibility of coupling two remote N-V centers via an intermediate chain of nitrogen spins.

The N-V center consists of a substitutional nitrogen atom next to a vacancy in the diamond lattice. We study here the negatively charged state (NV$^-$). It has a spin triplet ($^3$A) ground state, with a zero-field splitting $D$= 2.88~GHz between the sublevels with spin z-component $m_{S}^{NV}\!=\!0$ and $m_{S}^{NV}\!=\!\pm1$~\cite{GroundStates}. The arrangement of the triplet excited state ($^3$E) sublevels is less clear~\cite{ExcitedStates}. The spin is quantized along the N-V symmetry axis, a $<\!\! 111 \!\! >$ crystal axis~\cite{Loubser78}. Linearly polarized optical excitation preferentially pumps the spin system into the ground state $m_{S}^{NV}\!=\!0$ level~\cite{Harrison2004}. Also, the average photon emission rate is substantially smaller for transitions involving the $m_{S}^{NV}\!=\!\pm 1$ levels than for the $m_{S}^{NV}\!=\!0$ level~\cite{JelezkoAPL02}, which allows readout of the spin state by the photoluminescence intensity $I_{PL}$. These latter two effects have been attributed to spin-dependent intersystem crossing to a singlet ($^1$A) level~\cite{Loubser78,ExcitedStates,Nizovstev03}.

We study a set of single N-V centers in a single-crystal high-temperature high-pressure (type Ib) diamond. 
This diamond contains nitrogen (N) impurities with a density of $10^{19}-10^{20} $~cm$^{-3}$, corresponding to an average nearest-neighbour distance of a few nm. These N centers contain one excess electron carrying a spin of $\frac{1}{2}$. 
The average distance between nearest-neighbour N–V centres is about 3~$\mu$m, much larger than the spatial resolution of our setup ($\approx 0.3~\mu$m, set by the spot size of the laser beam). 

We detect the phonon-broadened $^3$E-$^3$A transition of the N-V center using non-resonant photoluminescence
microscopy (for details on the setup see Ref.~\cite{EpsteinNatPhys05}). Single N-V centers are identified through photon antibunching
measurements (establishing that it is a single emitter) and ESR measurements that reveal the characteristic zero-field splitting $D$. We precisely align the external magnetic field $B$ with the [111] crystal axis, which is the symmetry axis for all centers studied in this work and will be referred to as the $z$-axis. All experiments are performed at room temperature.

The relevant parts of the Hamiltonians $H_{NV}$ and $H_{N}$ of the N-V spin and N spin respectively, are 
\begin{eqnarray}
	H_{NV}&=&D (S_{z}^{NV})^2 + g_{NV} \mu_B B S_{z}^{NV},\\
	H_N&=&g_N \mu_B B S_{z}^{N} + A\ \vec{S}^{N}\! \cdot\! \vec{I}^{N},
\end{eqnarray}
where $g_{NV}$ and $g_N$ are the electron $g$-factors (2.00 for both centers), $\mu_B$ is the Bohr magneton, $\vec{S}^{NV} (\vec{S}^{N})$ is the electron spin operator of the N-V center (N center) and $\vec{I}^{N}$ is the nuclear spin operator of the nitrogen atom. Virtually all nitrogen atoms in diamond are $^{14}$N with total spin $I^N$=1. The nitrogen hyperfine (HF) constant $A$ depends on the orientation of the HF axis with respect to $B$. The HF axis is set by a Jahn-Teller distortion of the N atom towards one of its four neighbouring carbon atoms, resulting in four possible orientations. For $B$ aligning the electron spins along [111], three of these orientations are equivalent and have $A$=86~MHz, while the other one has $A$=114~MHz~\cite{SmithPR59}. The electrons at the N-V center also experience a HF interaction, but the resulting frequency splittings (about 2~MHz~\cite{Loubser78}) are not resolved in this work.

\begin{figure}[tbp]
	\includegraphics[width=3.4in]{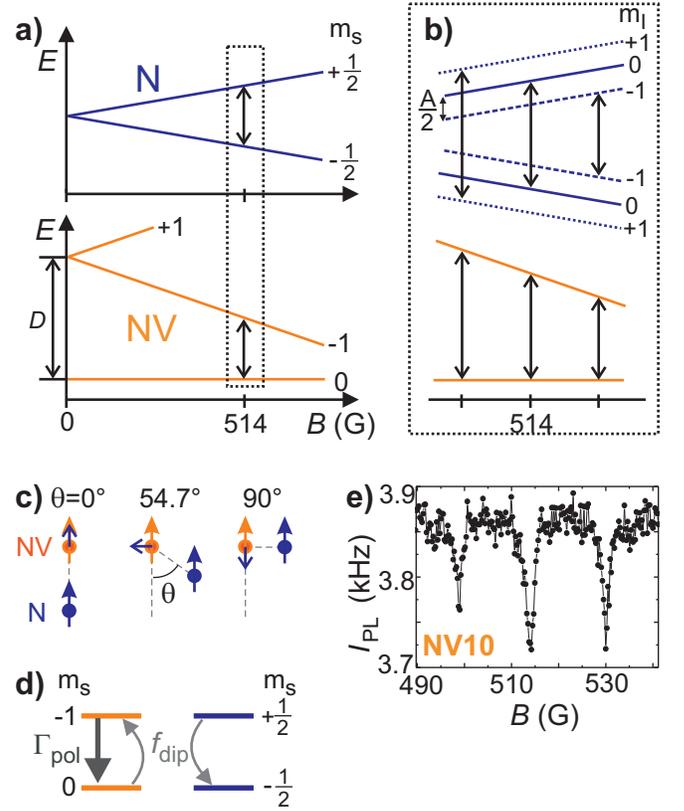}
	\caption{\textbf{\textbf{(a)}} Energy levels of the N-V and a N electron spin as a function of a magnetic field $B$.
\textbf{(b)} Zoom-in of (a) around 514~G, showing how the splitting of the N electron states into triplets due to the hyperfine interaction leads to three different energy resonances with the N-V center. \textbf{(c)} The dipolar field $\vec{B}_{\mathrm{dip}}$ (small blue arrow) induced by the N spin (blue dot-arrow) at the site of the N-V center (orange dot-arrow) for different orientations. \textbf{(d)} Schematic of the spin dynamics at resonance. The optically-induced spin polarization of the N-V center (with rate $\Gamma_{pol}$) is reduced by flip-flops with the N spin at a rate $f_{\mathrm{dip}}$. \textbf{(e)} Photoluminescence intensity $I_{PL}$ as a function of $B$ for NV10, showing clear dips due to the dipolar coupling at the resonant fields~\cite{laserESRpower}.}
	\label{fig:fig1}
\end{figure}

Figure~\ref{fig:fig1}a shows the resulting energy levels of the N and the N-V electron spins as a function of $B$. At $B$=514~G, the energy difference between the N doublet equals the energy difference between the $m_S^{NV}\!=\!0$ and $m_S^{NV}\!=\!-1$ sublevels of the N-V center. Due to the HF interaction of the nitrogen spins, this resonance condition has sidepeaks as indicated in Fig.~\ref{fig:fig1}b. In the rest of this work, we disregard the $m_{S}^{NV}\!=\!+1$ state, as it is far away in energy from the $m_{S}^{NV}\!=\!0$ and $m_{S}^{NV}\!=\!-1$ states and therefore has a negligible effect on their spin dynamics.

The nitrogen doublet is coupled to the N-V center spin via the magnetic dipolar coupling~\cite{vanOortPRB90}:
\begin{equation}
	H_{\mathrm{dip}}\!=\!\frac{\mu_0 g_{NV}g_N \mu_B^2}{4 \pi r^3} \left[\vec{S}^{NV}\!\!\cdot\!\vec{S}^{N}\!-\!3(\vec{S}^{NV}\!\!\cdot\!\hat{r})(\vec{S}^{N}\!\cdot\! \hat{r})\right],
	\label{eq:dipolar1}
\end{equation}
where $\mu_0$ is the magnetic permeability, $r$ is the distance between the centers and $\hat{r}$ the unit vector connecting them. The prefactor is about 6.5~MHz for $r$=2~nm.

Due to the dipolar coupling, the N-V electron spin feels a magnetic field $\vec{B}_{\mathrm{dip}}$ caused by the N electron spin, and vice versa. The effect of $\vec{B}_{\mathrm{dip}}$ depends crucially on the relative orientations of the spins, as can be seen from the three examples shown in Fig.~\ref{fig:fig1}c, where $\vec{B}_{\mathrm{dip}}$ at the N-V center is depicted as a blue arrow. We define $\theta$ as the angle between $\hat{r}$ and $\hat{z}$. 

For $\theta$=54.7$^\circ$, also known as the ``magic angle''~\cite{Slichter}, $\vec{B}_{\mathrm{dip}}$ is perpendicular to $\hat{z}$. Therefore, the two-spin states ($m_{S}^{NV},m_{S}^{N}$)=(0,+$\frac{1}{2}$) and (-1,-$\frac{1}{2}$) are strongly mixed by the dipolar fields at the resonance condition depicted in Fig.~\ref{fig:fig1}a. The other states are far away in energy and remain good eigenstates. Figure~\ref{fig:fig1}d sketches the resulting dynamics. Optical illumination polarizes the N-V center spin into $m_{S}^{NV}\!=\!0$ with a rate $\Gamma_{pol}$, whereas the component of $\vec{B}_{\mathrm{dip}}\perp \hat{z}$ leads to flip-flops between the N-V and N electron spins, at a rate $f_{\mathrm{dip}}$ determined by Eq.~\ref{eq:dipolar1}, that reduce the N-V center spin polarization. This process is revealed in magnetic field scans as shown in Fig.~\ref{fig:fig1}e~\cite{EpsteinNatPhys05}. The photoluminescence intensity $I_{PL}$, which is proportional to the N-V center spin polarization, dips at the resonance conditions shown in Fig.~\ref{fig:fig1}b, indicating loss of spin polarization. 

\begin{figure}[t]
	\includegraphics[width=3.4in]{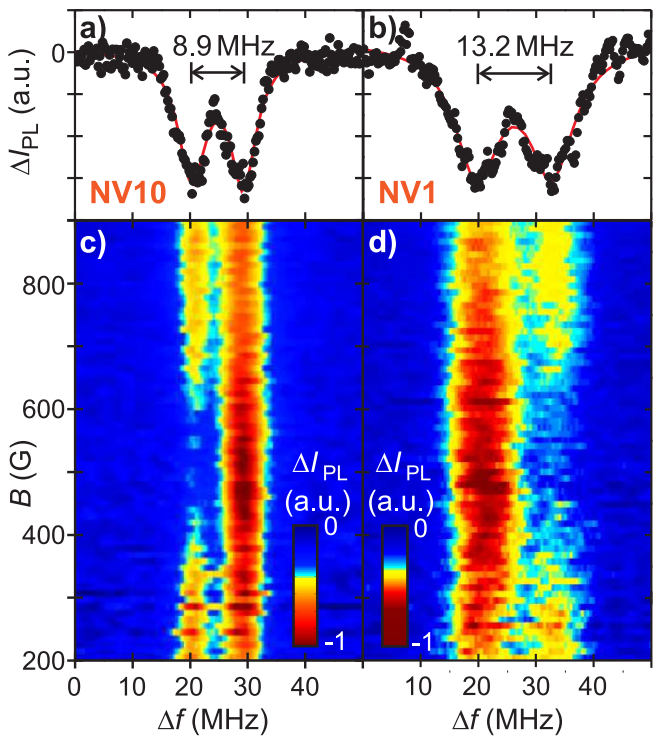}
	\caption{Room-temperature coupling of the N-V center to a single N spin~\cite{laserESRpower}. \textbf{(a)-(b)} Electron spin resonance on NV10 (a) and NV1 (b) at $B$=100~G. The ESR-induced change in $I_{PL}$, $\Delta I_{PL}$,  is plotted versus ESR frequency around the resonance frequency $(D-g_{NV}\mu_B B)/h$. The resonance is split into a doublet, whose components correspond to the two possible orientations of a single nearby N spin. \textbf{(c)-(d)} Magnetic field dependence of the ESR-induced $\Delta I_{PL}$ for NV10 (c) and NV1(d).} 
	\label{fig:fig2}
\end{figure}
For $\theta$=0, $\vec{B}_{\mathrm{dip}}$ is parallel to $\hat{z}$, and the magnitude of total magnetic field felt by the N-V electron spin depends on the state of the N spin. As a result, the state ($m_{S}^{NV},m_{S}^{N}$)=(-1,-$\frac{1}{2}$) is lower in energy than (-1,+$\frac{1}{2}$). However, the \textit{direction} of the total field is along $\hat{z}$ as in the uncoupled case and therefore the eigenstates are not changed. The same is true for $\theta$=90$^\circ$, but in this case the sign of the dipolar field is reversed such that (-1,+$\frac{1}{2}$) is lower in energy than (-1,-$\frac{1}{2}$). Note that for both $\theta$=0 and $\theta$=90$^\circ$, the energy of the $m_{S}^{NV}$=0 state is not affected by the dipolar coupling.

Thus, if the N-V center is strongly coupled to only a single N electron spin, the $m_{S}^{NV}$=-1 state is split into a doublet by the component of $\vec{B}_{\mathrm{dip}}\ /\!/\ \hat{z}$, with the two components corresponding to the two possible orientations of the N electron spin. In case of strong coupling to more than one spin, the total number of possible orientations of these spins is more than 2 and therefore the $m_{S}^{NV}$=-1 state will be split into a multiplet. 

Figures~\ref{fig:fig2}a-b show ESR between the $m_{S}^{NV}$=0 and -1 states at $B$~=~100~G for two different centers (NV10 and NV1). When the applied frequency is resonant with the $m_{S}^{NV}=0\!\leftrightarrow\!-1$ transition, the optically induced spin polarization is reduced and $I_{PL}$ drops. A clear splitting of this resonance in two is observed, indicating that these centers are each predominantly coupled to a \textit{single} N spin. Since the ESR frequency depends on the state of the N electron spin, this state can be read out by monitoring the presence or absence of ESR-induced $\Delta I_{PL}$ at one of the two ESR frequencies. At three other centers we found the splitting of the resonance to be smaller than the linewidth (a few MHz).

By repeating ESR scans at different magnetic fields and plotting the ESR-induced $\Delta I_{PL}$ in color scale, we obtain a 2D plot as depicted in Fig.~\ref{fig:fig2}c for NV10. Strikingly, one of the two ESR-induced dips disappears completely around 500~G, while the other dip gaines in amplitude. This indicates that at these magnetic fields the N electron spin is strongly polarized~\cite{WrachtrupCoupling}. From our previous analysis (see Fig.~\ref{fig:fig1}d) it follows that the N spin must be polarized into $m_{S}^{N}$=-$\frac{1}{2}$ state, and the disappearing dip therefore corresponds to the $m_{S}^{N}$=+$\frac{1}{2}$ state. Since the resonance for $m_{S}^{N}$=+$\frac{1}{2}$ is found at the lower of the two frequencies, the coupled state ($m_{S}^{NV},m_{S}^{N}$)=(-1,+$\frac{1}{2}$) has a lower energy than the (-1,-$\frac{1}{2}$) state. Therefore, the dipolar coupling is antiferromagnetic for this pair and the angle $\theta$ must be near 90$^\circ$.

\begin{figure}[tbp]
	\includegraphics[width=3.4in]{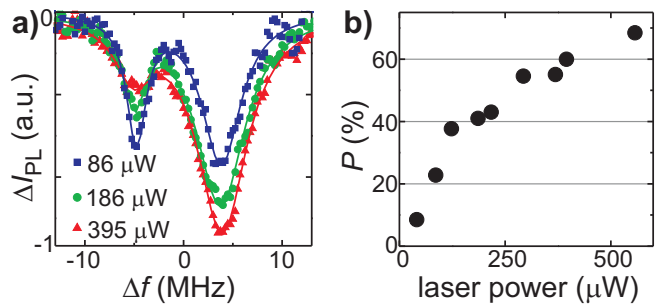}
	\caption{Dependence of the N electron spin polarization on laser power at $B$=600~G for NV10~\cite{laserESRpower}.
	\textbf{(a)} $\Delta I_{PL}$ versus ESR frequency, for three different laser powers.
	\textbf{(b)} Polarization $P$ of the N spin as a function of laser power.}
	\label{fig:fig3}
\end{figure}
The result of similar measurements on NV1 is shown in Fig.~\ref{fig:fig2}d. Again, we observe strong polarization of the N spin, but in this case the high-frequency dip disappears. From similar reasoning as above we conclude that the coupling for this pair is ferromagnetic, and $\theta$ must be closer to 0$^\circ$. Knowing the value of the z-component (half the splitting between the ESR dips) and the sign of the dipolar coupling, we deduce an upper bound on the distance between the N atom and the N-V center using Eq.~\ref{eq:dipolar1} of 2.3~nm for NV10 and 2.6~nm for NV1.

The magnetic field range over which polarization of the nitrogen spin occurs is related to the laser power. Higher laser power leads to a stronger polarization rate of the N-V center $\Gamma_{pol}$ (see Fig.~\ref{fig:fig1}d). Since this N-V center polarization rate is a source of decoherence for the coupled-spin system, higher laser powers lead to more broadening of the dipolar resonance.

We investigate the laser dependence of the N spin polarization in more detail at $B$=600~G. Figure~\ref{fig:fig3}a shows ESR at NV10 for three different laser powers. An evolution is observed from almost equal populations of the two N spin orientations for low laser power to a high polarization as the laser power is increased. In Fig.~\ref{fig:fig3}b we plot the polarization $P$ as a function of laser power. Here,
\begin{equation}
	P=\frac{\Delta I_{PL}[-\frac{1}{2}]-\Delta I_{PL}[+\frac{1}{2}]}{\Delta I_{PL}[-\frac{1}{2}]+\Delta I_{PL}[+\frac{1}{2}]},
	\label{eq:P}
\end{equation}
with $\Delta I_{PL}$[-$\frac{1}{2}$] ($\Delta I_{PL}$[+$\frac{1}{2}$]) the ESR-induced $\Delta I_{PL}$ at the resonance frequency for $m_{S}^{N}\!=\!-\frac{1}{2}$ ($m_{S}^{N}\!=\!+\frac{1}{2}$). Because the two dips have some overlap, we fit the curves with a double Lorentzian and obtain $\Delta I_{PL}$ from the amplitude of the corresponding single Lorentzian. We find that the $P$ rises from close to zero to about 70\% for a laser power of 550~$\mu$W. For even higher laser powers the polarization seems to exceed 75\%, but in this regime the strong overlap of the two ESR-induced dips prevents a reliable fit to the data. The saturation of $P$ at high laser power arises from the balance between the polarization and spin relaxation processes.

\begin{figure}[t]
\includegraphics[width=3.2in]{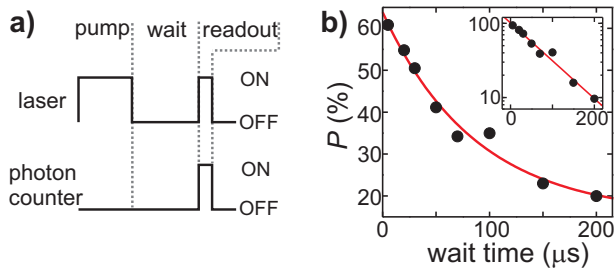}
\caption{Measurement of the relaxation time of a single nitrogen spin. \textbf{(a)} The pump-probe sequence as used in the experiments. The pump time is 100~$\mu$s and the readout time is 5~$\mu$s. A pattern generator controls the laser via an Acoustic-Optical Modulator, and the photon counter by gating it. Laser power is 395~$\mu$W.
\textbf{(b)} Polarization $P$ of a single nitrogen spin as a function of wait time~\cite{laserESRpower}. The red line is a single-exponential fit to the data, with a  $T_1$ of 75~$\mu$s.
Inset: same data  on a logarithmic scale, with the offset subtracted and the decaying part of the data normalized to 100\%.
}
\label{fig:fig4}
\end{figure}

The results presented in Figs.~\ref{fig:fig2} and ~\ref{fig:fig3} demonstrate our ability to polarize a single N electron spin and to read out its state. We now apply these methods to probe the lifetime of this spin at $B$=600~G, using the pump-probe scheme shown in Fig.~\ref{fig:fig4}a. We first apply a 100~$\mu$s laser pulse that polarizes the spin. Then the laser is turned off for a variable wait time, allowing the spin to relax to thermal equilibrium. After the wait time, a short laser pulse of 5~$\mu$s is applied during which the photon counter is turned on to monitor $I_{PL}$. Finally, both the laser and the photon counter are turned off for a time that compensates the wait time in such a way that the total cycle time is constant. This pump-probe cycle is repeated about 2000 times per second while we scan across the two ESR frequencies. From the resulting curve, the polarization is determined as before.

In Figs.~\ref{fig:fig4}b we plot $P$ as a function of wait time. A clear decay of the spin polarization is observed. The value close to zero wait time is the same as for continuous laser illumination with the same power (see Fig.~\ref{fig:fig3}b). Since these experiments are performed at room temperature where $P\approx$0 in thermal equilibrium, one would expect $P$ to drop all the way to zero for longer wait times. However, during the readout pulse the polarization is slowly building up again, resulting in a finite polarization at the end of the readout pulse even if it was zero at the start.

The data seem to follow a single-exponential decay law (see inset of Fig.~\ref{fig:fig4}b). From a fit to the data we find a spin relaxation time of $T_1\!=\!(75\pm20$)~$\mu$s for this single nitrogen spin at room temperature. This is an order of magnitude shorter than spin-lattice relaxation times found on an ensemble~\cite{Reynhardt98}, indicating that spin-spin interactions might play a dominant role in single spin relaxation.

Finally, we briefly return to the data of Fig.~\ref{fig:fig1}e. A detailed look reveals that the dips corresponding to $m_I^N$=+1 (near 499~G) and $m_I^N$=-1 (near 530~G), which are equivalent~\cite{SmithPR59}, have different amplitudes. This could indicate that the nuclear spins are not in equilibrium, but are rather polarized into the $m_I^N$=-1 state. This can be the result of hyperfine-induced flip-flops between the polarized N electron spin and the nuclear spin. This spin trade takes the electron from the $m_S^N$=-$\frac{1}{2}$ state (into which it is polarized by the dipolar coupling) to the $m_{S}^{N}$=+$\frac{1}{2}$ state, while at the same time decreasing the nuclear spin z-component by 1. The result of many such events is a net polarization of the nuclear spins in the $m_{I}^{N}$=-1 state. This nuclear polarization will be subject of future work.

In summary, we have observed coupling of an N-V center in diamond to a single nitrogen spin in the surrounding lattice at room temperature. By using the laser as a pump and ESR measurements as a probe, we have demonstrated strong polarization of a single nitrogen electron spin, and we have determined its spin relaxation time through time-resolved pump-probe measurements. Finally, we have found indications that the nuclear spin of the nitrogen atom is also polarized through the hyperfine interaction. These results constitute a significant step towards room-temperature quantum information processing in diamond.

This work was supported by AFOSR, DARPA/MARCO, DARPA/CNID and ARO.


\begin{thebibliography}{99}
\bibitem{laserESRpower}
Laser power is 186~$\mu$W for Figs.~\ref{fig:fig1}e and \ref{fig:fig2}. The ESR output power at the source is 10~dBm in Figs.~\ref{fig:fig2}-\ref{fig:fig3} and 12~dBm in Fig.~\ref{fig:fig4}.
\end{thebibliography}
\end{document}